\documentstyle[epsfig,twoside,rotating]{aa}
\newcommand{\Msol}{M_\odot}

\begin{document}
\title{Study of a Strategy for Parallax Microlensing Detection Towards
the Magellanic Clouds}
\author{
S.~Rahvar\inst{1,2},
M.~Moniez\inst{3},
R.~Ansari\inst{3},
O.~Perdereau\inst{3}
}
\institute{
Department of Physics, Sharif University of Technology,
P.O. Box 11365-9161, Tehran, Iran
\and
Institute for Studies in Theoretical Physics and Mathematics,
P.O. Box 19395-5531, Tehran, Iran
\and
Laboratoire de l'Acc\'{e}l\'{e}rateur Lin\'{e}aire,
{\sc IN2P3-CNRS}, Universit\'e de Paris-Sud, B.P. 34, 91898 Orsay Cedex, France
}

\offprints{M. Moniez: moniez@lal.in2p3.fr.}

\date{Received ??/??/1999, accepted }

%\thesaurus{10.08.1;10.11.1;10.19.2;10.19.3;12.04.1;12.07.1}

%%%%%%%%%%%%%%%%%%%%%%%%%%%%%%%%%%%%%%%%%%%%%%%%%%
%                                                %
%    BEGINNING OF TEXT                           %
%                                                %
%%%%%%%%%%%%%%%%%%%%%%%%%%%%%%%%%%%%%%%%%%%%%%%%%%
%\maketitle
%\tableofcontents

\abstract{
In this article, we have investigated the possibility to distinguish
between different
galactic models through the microlensing parallax studies.
We show that a systematic search for
parallax effects can be done using the currently running alert systems and
complementary photometric telescopes, to distinguish between
different lens distance distributions.
We have considered two galactic dark compact objects distributions,
with total optical depths corresponding to the EROS current upper limits.
These models correspond to two extreme hypothesis on a three component
galactic structure made of a thin disk, a thick disk, and a spherically
symmetric halo.
Our study shows that for sub-solar mass lenses, an exposure of $8\times 10^7$
star$\times$yr toward LMC should allow to distinguish between these two
extreme models.
In addition the
self-lensing hypothesis (lensing by LMC objects) can
efficiently be tested through the method proposed here.

}

\maketitle

\markboth{S. Rahvar et al. : Parallax Microlensing Detection Towards
the Magellanic Clouds}{S. Rahvar et al. :
Parallax Microlensing Detection Towards
the Magellanic Clouds}

\section{Introduction}               

Following the suggestion of Paczy\'nski (\cite{pacz}), several microlensing
surveys are underway to search for dark matter in the form of Massive 
Compact Halo Objects (MACHO's).
Towards the Magellanic Clouds, 
13 to 17 microlensing candidates (depending on the quality)
have been found by the MACHO
collaboration (\cite{macholmc}), 3 candidates were found by the
EROS team towards the {\it LMC} (\cite{notenoughmachos}) and one 
towards the {\it SMC} (\cite{palanque}). The OGLE collaboration 
(\cite{udalski}) also reported one event towards the {\it LMC}.\\
One of the critical issues concerning the galactic dark matter problem
is the localization of the deflecting objects.
One way
to obtain more information about the location of the lenses
is to measure the shape of the light curve with a precision
allowing the detection of second order effects.
Amongst those effects, the possible use of the well controlled
Parallax effect was discussed by (\cite{grieger}) 
and (\cite{gould}).
Parallax effect has also been studied towards 
the galactic bulge by (\cite{buchalter}) and {\it LMC} by 
(\cite{gould98}).
Few parallax microlensing has been observed
up to now (\cite{machopar}; \cite{mao}; \cite{Soszy}). However the non-observation
of the effect has sometimes been used to put constraints on
the parameters of the lens (\cite{palanque}, \cite{BS2ans}). 
In order to increase the sensitivity to the parallax effect,
an accurate photometry and a high sampling rate 
are necessary. 
One of the ways to reach this aim is to exploit the alert systems of the
microlensing surveys with a follow-up setup.
EROS is one of the groups that is using an alert system to trigger 
ongoing microlensing events.  
The aim of the work presented here is to simulate the observation
of microlensing alerts triggered by EROS
with a follow-up telescope, and to estimate the efficiency of such
a combination to detect the parallax effects.
In Section 2, we review
the characteristics of parallax distorted events. In Section 3, we
present the strategy implemented in our simulation, according 
to realistic conditions of observation, and we discuss
the fitting procedures to the light curves.
In Section 4, we give the efficiency of parallax detection
and the number of expected parallax events for two extreme
galactic models. These results are then discussed in section 5.
%************************************************************************
\section{Parallax Microlensing Events}
\subsection{Simple Microlensing}
So-called standard or simple microlensing events occur when the approximation
of a point-like deflector and point-like source with a relative uniform
motion is valid.
At a given time $t$,
the light magnification $A(t)$ of a pointlike source located at distance $D_s$
induced by a pointlike deflector of mass $M$ located at distance $xD_s$
is given by:
\begin {equation}
A(t) = \frac{ {u(t)}^2 + 2} {u(t)\sqrt{4 + {u(t)}^2}}
\end {equation}
where $u(t)$ is the distance between the undeflected
line of sight and the deflecting object, expressed
in units of the ``Einstein Radius" $R_E$,
the characteristic length of the phenomenon:
\begin{eqnarray}
 & R_E & = \sqrt{ \frac{4GM}{c^2}D_{s}x(1-x)},\\
&\simeq&\ 4.5\ A.U. \times\left[\frac{M_D}{\Msol}\right]^{\frac{1}{2}}
\times\left[\frac{D_s}{10\ Kpc}\right]^{\frac{1}{2}}
\times\frac{\left[x(1-x)\right]^{\frac{1}{2}}}{0.5}.\nonumber
\end{eqnarray}
Here $G$ is the Newtonian gravitational constant.
Assuming a deflector moving at a constant relative transverse
speed $v_T$, reaching its minimum
distance (minimum impact parameter) to the undeflected line
of sight $u_0$ at time $t_0$, $u(t)$ is given by
$$ u(t)= \sqrt{{u_0}^2 + (\frac{t - t_0}{t_E})^2},\ where\  
t_E  = \frac{R_E}{v_T}$$ the ``lensing time scale" is
the only measurable parameter
bringing useful information on the deflector in the approximation
of the simple microlensing.
Within this approximation,
the light curve of a microlensed star with base flux $F_b$
is fully determined by the 4 parameters
$ F_b, u_0, t_0, t_E$. \\

\subsection{Parallax Effect in Microlensing}
\begin{figure*}
\begin{center}
\mbox{\epsfig{file=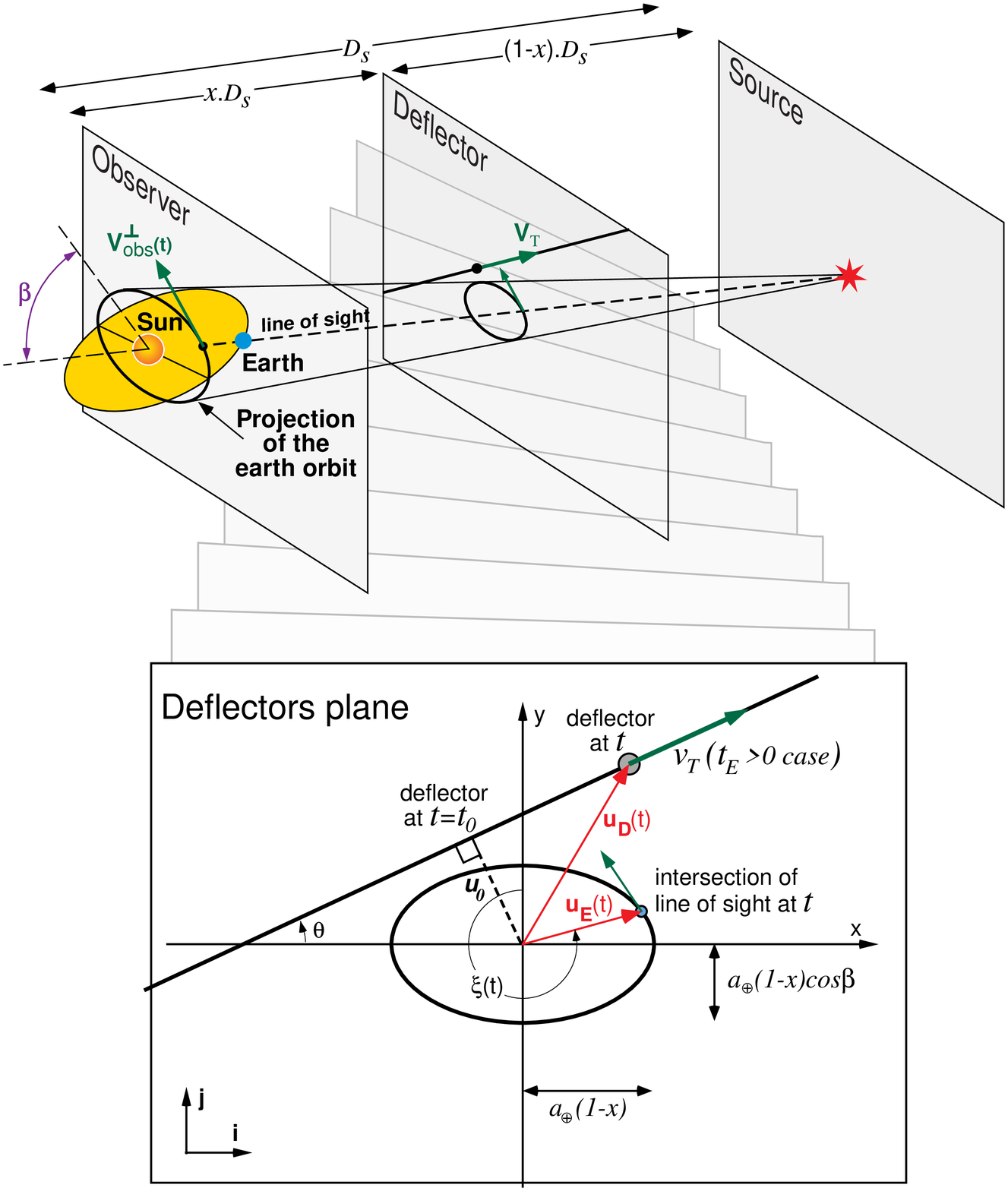,width=14.cm}}
\caption[]{Principle of the parallax effect, and notations used
in the text.
Here, the lensing configuration is drawn in a referential where
the Sun-source line is at rest.
Note that changing the sign of $t_E$ changes
the direction of the deflectors speed. This would have no
impact if parallax is neglected.
\label{parallax}}
\end{center}
\end{figure*}

If the variation of the Earth's velocity rotating component
around the sun is
not negligible with respect to the projected transverse speed of the
deflector, then the apparent trajectory of the deflector with
respect to the line of sight is a cycloid instead of a straight line.
The resulting amplification versus time curve is therefore
affected by this parallax effect. This effect is more easily
observable for long duration events (several months), for
which the change in the Earth velocity is important (\cite{machopar}).
If ${\bf  u_D(t)}$ is the position of the deflector in the deflector's
transverse plane and ${\bf u_E(t)}$ is the intercept of the Earth-source
line of sight with this plane (see Fig. \ref{parallax}) , then
\begin {equation}
\label{u}
{u(t)}^2 = |{\bf  u_D(t)} - {\bf u_E(t)}|^2.
\end {equation}
${\bf u_D(t)}$ is given by:
\begin{eqnarray}
{\bf u_D(t)}&=& (\frac{ t -t_0}{t_E}cos(\theta) - 
u_0 sin(\theta)) {\bf i} \nonumber\\
 &+& (\frac{ t -t_0}{t_E}sin(\theta) + 
u_0cos(\theta)) {\bf j}, 
\end{eqnarray}
where $\theta$ is the angle between the projected lens trajectory
and the projected major axis
of the Earth orbit in the deflector's plane, $t_0$ is the time 
of the closest approach $u_0$ of the lens to the sun-source
line of sight, and $t_E$ is the Einstein radius crossing time
when neglecting the Earth motion (see Fig. \ref{parallax}).
Here we have chosen the convention to consider negative values of $t_E$,
which mathematically describe the configurations where
the direction of the deflector's kinetic momentum
${\bf u_D(t)}\wedge {\bf v_T}$ is inverted,
in order to span every possible configuration\footnote{
When parallax is not taken into account, the light curves
obtained after inversion of the direction of the velocity are
the same, this is not anymore the case with parallax.}.
Neglecting the earth orbital excentricity, ${\bf u_E(t)}$ is given by: 
\begin{equation}
{\bf u_E(t)}  =  \delta u\ sin(\xi(t)){\bf i } + \delta u\ cos(\xi(t)) cos(\beta)
{\bf j}, 
\end{equation}
where $\delta u=a_{\oplus}(1-x)/R_E$
is the projection of the earth orbital radius in the deflector
plane in unit of Einstein radius, $\beta$ is the angle between 
the ecliptic and deflector plane and $ \xi(t)$ is the phase of the earth
relative to its position when ${\bf u_E} = \delta u\ cos(\beta) {\bf j}$. 
The distortion of the light curve
is important if the Earth's orbital velocity around the sun is
not negligible with respect to the projected transverse speed of the
deflector
$$\tilde v=\frac{R_E}{t_E (1-x)}=\frac{a_{\oplus}}{\delta u.t_E}.$$
\section{Simulation of a systematic search for parallax}
To study the response of the proposed combination of a trigger system
with dedicated follow-up capabilities,
we have developed a Monte-Carlo simulation of events with parallax
effects.
In this section, we describe the generation of the event parameters,
the EROS trigger system response and the simulation
of the expected highly sampled light-curves provided by the follow-up
telescope. We then describe a fitting procedure from which we
estimate the parallax and its significance for each event.
Detection efficiency functions are obtained, which can be combined with
any spatial distribution of lenses.
\subsection{Simulation of the microlensing events}
As mentioned above, simple microlensing light curve depends on 4
parameters, namely the base flux $F_b$,
$u_0$, $t_E$ and $t_0$.
To take into account the parallax, two extra parameters
$\theta$ and $\delta u$ need to be generated.
The direction of the target, on which depend
$\xi(t_0)$ and $\beta$ should also be fixed.
To calculate an efficiency function with uniform statistical
precision we generate uniformly the following parameters in their
respective intervals:
$u_0\in[0,2]$, $\log(|t_E\,\mathrm{(days)}|)\in[0.7,2.6]$
(positive and negative values for $t_E $), $t_0\in[t_{first}-50,
t_{last}+50]\,\mathrm{(days)}$ where $t_{first} = 50$ and $t_{last} = 750$,
$\theta\in[0,2\pi]$ and $\delta u\in[0,10]$.
\subsection{Simulation of EROS-like light curves}
The generated parameters of a microlensing event together with the
coordinates of the lensed object allows us to write
the analytical expression of the magnification versus the time.
We keep only the events with $A_{Max}>1.34$\footnote{Notice that
this is not equivalent to select events with $u_0<1$ due to the parallax.}.
The base fluxes $F_b$ of the lensed stars are chosen according
to the observed distributions in the EROS catalogues towards
{\it LMC} and {\it SMC}.
Light curves are simulated by using the effective sampling
rate of EROS towards {\it LMC} and {\it SMC} (about one observation per six
nights in average).
Every measurement is then randomly shifted according to a
Gaussian distribution that reflects the photometric uncertainties.
The average relative photometric precision $\Delta F/F$ for a given
flux $F$ (in ADU unit)
is taken from the EROS phenomenological parametrisation
found for a standard quality image (\cite{TheseFred}):
\begin {equation} 
\frac{\Delta F}{F} = 3.5\times F^{-0.85},
\end{equation} 
\subsection{Simulation of a simple alert system}
\begin{figure}
\begin{center}
\begin{turn}{0}
\mbox{\epsfig{file=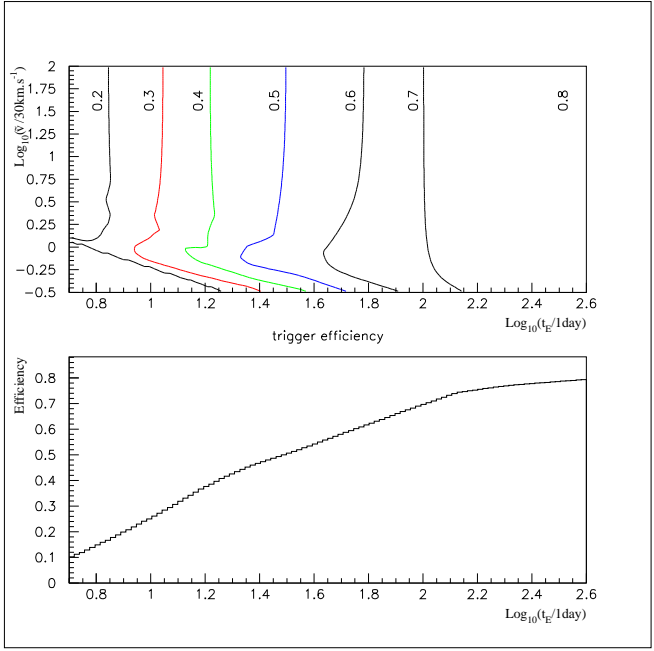,width=8.6cm}}
\end{turn}
\caption[]{The efficiency of the alert procedure
in the $Log_{10}(\tilde v/30\,\mathrm{km.s^{-1}})$ vs $Log_{10}(t_E/1.\,\mathrm{day})$
plane, for 700 days of observation towards the Magellanic Clouds.
This efficiency is the ratio of the number of events
that satisfy the trigger
conditions, to the number of events -- generated with any $u_0$
and any date of peak magnification
within the $700\pm 50$ days period -- with
maximum magnification $A_{max}>1.34$.
The average efficiency as a function of
$Log_{10}(t_E/1.\,\mathrm{day})$ is also shown.
At this stage, $Log_{10}(\tilde v/30\,\mathrm{km.s^{-1}})$ cannot always been
experimentally measured.
\label{effictrig}}
\end{center}
\end{figure}

The next step is to simulate an alert system for ongoing events,
which is necessary to trigger follow-up observations.
According to one of the EROS alert algorithms,%as follows:
we consider that we will monitor events as soon as their
light curves exhibit
4 consecutive flux measurements above 4 standard deviations
from the base line (\cite{mansoux}). 
It is clear that only the most significant microlensing events are
selected by this algorithm.
We have in fact considered several trigger thresholds, from a
loose criterion (3 consecutive measurements above $3\sigma$ from the
base line) to the strict criterion we finally use. Even using this
strict criterion, an average of
one false alarm due to variable stars or instrumental
artifacts is expected per true microlensing alert (\cite{JFGprivate}).
This false alarm rate will induce some wasted follow-up time, but
for very limited durations, as non-microlensing events would be
identified quickly.
Figure \ref{effictrig} shows the trigger efficiency as a function of
the parallax parameter $Log_{10}(\tilde v/30\,\mathrm{km.s^{-1}})$ and the event duration
$Log_{10}(t_E/1.\,\mathrm{day})$, averaged over
all the other parameters. This efficiency is relative to the events
for which the earth enters the projected Einstein radius $R_E$, i.e.
for events with maximum magnification $A_{max}$ larger than 1.34.
Note that in the case of strong parallax effect, due to the non-linear
trajectory of the earth,
this definition may considerably differ from the usual
definition of the efficiency relative to events with impact
parameter $u_0<1$.
More details about the efficiency of this algorithm can be found
in (\cite{rahvar}).
 
We assume that the subsequent monitoring
of the on-going microlensing events will be made with a
telescope large enough to achieve a 1\% precision photometry.
We also assume that the light curves will be measured
two times every night during which the observability of the source
exceeds 3 hours, and one time during the other nights.
The probability of good weather is taken as $70\%$.
We simulate light curves accordingly to these conditions for one colour.
Figure \ref{lightcurve} shows an example of the simulated light curve
including the pre-alert EROS-like measurements and the more precise
post-alert measurements.
\subsection
{Fitting Procedure}
\begin{figure}[h!]
\begin{center}
\begin{turn}{0}
\mbox{\epsfig{file=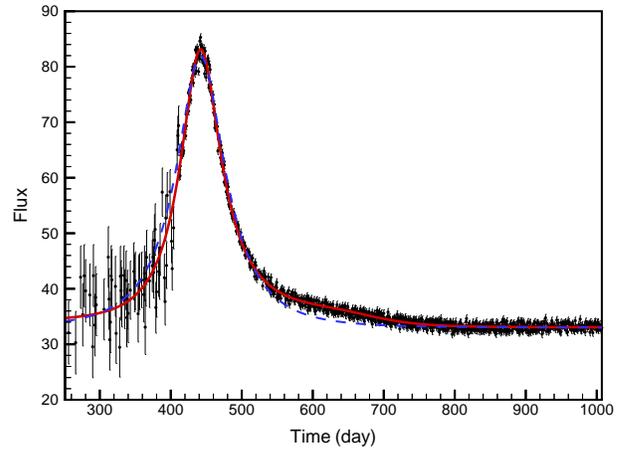,width=8.5cm}}
\end{turn}
\caption[]{Simulated long timescale microlensing event
with $2\%$ photometric precision (flux is in arbitrary units).
Pre-alert measurements are affected by much larger error bars
(about 10\%).
The full line corresponds to the parallax fit ($\chi^2 = 1318./(994\,\mathrm{d.o.f})$)
and the dashed line corresponds to the Standard fit
($\chi^2 = 4068/(996\,\mathrm{d.o.f})$). 
\label{lightcurve}}
\end{center}
\end{figure}

After simulating an event that has triggered follow-up observations,
we want to estimate the sensitivity to the parallax effect.
To this aim, we use
a $\chi^2$ minimization procedure to reconstruct the 6 microlensing
parameters with their errors.
Figure \ref{lightcurve} allows one to compare the results of this
parallax fitting procedure (6 parameters)
with a fit that neglects the parallax (4 parameters).
In this spectacular example, the $\chi^2$ of the best parallax
fit is smaller by
2750 units than the $\chi^2$ of the best non-parallax fit (for only
2 extra parameters).
In the following, we
will consider the significance of a parallax signal in units of
standard deviation from zero.
\section{Detection of parallax events}
\begin{figure}
\begin{center}
\begin{turn}{0}
\mbox{\epsfig{file=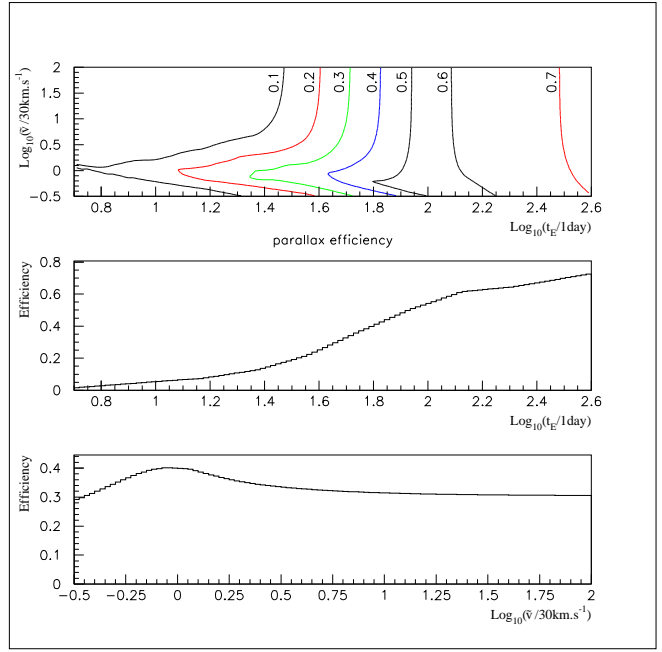,width=8.6cm}}
\end{turn}
\caption[]{Efficiency of the parallax detection
towards the Magellanic Clouds, as a function of
$Log_{10}(\tilde v/30\,\mathrm{km.s^{-1}})$ vs $Log_{10}(t_E/1.\,\mathrm{day})$ (top),
as a function of $Log_{10}(t_E/1.\,\mathrm{day})$ averaged on $\tilde v$ (middle),
and as a function of $Log_{10}(\tilde v/30\,\mathrm{km.s^{-1}})$,
averaged on $t_E$ (bottom).
Here we assume a photometric follow-up
precision of $1\%$ after the alert detection.
The efficiency is the ratio of the number of simulated events
satisfying the parallax detection conditions,
to the number of events generated with any $u_0$ and
any date of peak magnification within the $700 \pm 50$ days period,
and with $A_{max}>1.34$.
}
\label{effic}
\end{center}
\end{figure}

The probability to detect a parallax effect not only depends on
the 6 parameters that describe a microlensing light curve, but
also on the direction of the lensed source as mentioned in sect. 3.1.
In practice, we found that the relative change of the average
efficiencies from SMC to LMC can be neglected.
Figure \ref{effic} shows the efficiencies for detecting a parallax
effect towards the LMC
with $3\sigma$ signification in the $Log_{10}(\tilde v/30\,\mathrm{km.s^{-1}})$ versus
$Log_{10}(t_E/1.\,\mathrm{day})$
plane; these efficiencies are averaged over the parameters
that do not depend on the lens population, i.e. the source luminosity,
the impact parameter, the date of maximum and the orientation of the
lens velocity (assumed to be uniformly distributed from 0. to $2\pi$).
This 2-dimensional efficiency function can then be used to estimate numbers of
events from any model of the lens population
that predicts a $Log_{10}(\tilde v/30\,\mathrm{km.s^{-1}})$ versus
$Log_{10}(t_E/1.\,\mathrm{day})$ distribution.

We have checked that the ratio between significant parallax events
and triggered events is almost independent of the trigger threshold.
This will allow us to use the concept of relative efficiency to
detect parallax events with respect to triggered events $\epsilon_{par}$.
We thus assume that the global efficiency $\epsilon_{glob}$
can be written as the product of two terms:
$$\epsilon_{glob}=\epsilon_{trig}\times \epsilon_{par}.$$
Then, in the future, we will be able to obtain realistic numbers
by using this
relative efficiency with observed rates of events that relates to
an effective trigger.
As expected, the efficiency to detect a parallax event increases with
its duration.
\subsection{Parallax detection and Galactic Models}
We have produced
theoretical $Log_{10}(\tilde v/30\,\mathrm{km.s^{-1}})$ versus $Log_{10}(t_E/1.\,\mathrm{day})$ distributions
of microlensing events towards the ${\it LMC}$ and ${\it SMC}$,
for two extreme models.
The first model (model 1) has a ``thin'' disk
completely made of compact objects, with a standard isotropic
and isothermal halo including a maximum fraction of compact objects
that is compatible with the published EROS limits (\cite{notenoughmachos}).
The second model (model 2) has a ``thin'' plus a ``thick'' disks
both completely made of compact objects, and
a very light halo including a maximum fraction of compact objects
given by the EROS limits published in (\cite{theseLasserre}).
\begin{itemize}
\item{Halos}

The maximum contribution of compact objects in a halo
-- compatible with the EROS data -- depends on its mass function;
we consider here 4 different halo mass functions, namely Dirac distributions
peaked at $10^{-2}$,$10^{-1}$,$1.$ and $10.M_{\odot}$ respectively.
The corresponding maximum fractions of compact objects,
that we use in our simulation, are given in table 1 for the two haloes
considered here.
\begin{table}
\begin{center}
\caption[]{Maximum relative contribution of deflectors to the standard and
light halos (from \cite{notenoughmachos} \& \cite{theseLasserre}.}
\begin{tabular}{|l|c|c|}
\hline
			& model 1	& model 2 \\
Mass of deflectors	& standard halo	& light halo \\ \hline
$10^{-2} \Msol$		& 13\%		& 23\%	\\
$10^{-1} \Msol$		& 20\%		& 37\%	\\
$1. \Msol$		& 36\%		& 87\%	\\
$10. \Msol$		& 80\%		& 100\%	\\ \hline
\end{tabular}
\end{center}
\label{model}
\end{table}

The so-called standard Halo model has a spherical density distribution
$\rho(R)$ given by \cite{caldwell}:
\begin {equation}
\rho_H(R) = \rho_{H\odot}\frac{{{R_c}^2 +  R_{\odot}}^2}{R^2+{R_c}^2},
\end {equation}
where $\rho_{H\odot}$ is the local halo density,
$R_c$ is the halo ``core radius''
and $R_{\odot} = 8.5\,\mathrm{kpc}$    
is the Sun distance from the Galactic Centre.

The light halo model has a density distribution of a spherical
Evans type halo (\cite{evans94}) given by:
\begin{equation}
\rho_H(R) = \frac{V_a^2}{4\pi G}\frac{R^2+3R_c^2}{(R^2+R_c^2)^2}
\end{equation}
where the asymptotic velocity $V_a=170\,\mathrm{km/s}$ and G is the Newtonian
gravitational constant.
Velocities of halo objects follow the
Maxwell-Boltzmann distribution with a dispersion of $\sim 200\,\mathrm{km/s}$.
The observer and source motions can be neglected when considering
the transverse velocity with respect to the line of sight.

\item{Disks}

The mass functions of disks populations are taken from \cite{diskmassfunction}
for the two disks.
The density distribution in a disk is modeled
in cylindrical coordinates by a double exponential:
\begin{eqnarray*}
\rho_{D}(r,z) = \frac{\Sigma}{2H} \exp 
\left(\frac{-(r-R_{\odot})}{R_d} \right) \exp 
\left( \frac{-|z|}{H} \right) \ ,
\end{eqnarray*}
where $\Sigma$ is the column density of the disk at the Sun position,
$H$ the height scale and $R_d$ the length scale of the disk.
The distribution of the lens transverse velocity with respect
to the line of sight is established from the sun's proper motion
within the thin disk and from the local 
velocity distributions of the disk objects described by the
dispersion ellipsoids with
$\sigma_{r},\ \sigma_{\theta},\ \sigma_{z}$ dispersions;
the eventuality of a significant vertical gradient for the thick
disk rotation velocity is still a matter of debate.
At $1\,\mathrm{kpc}$ from the Galactic plane (mean distance of the thick-disk lenses),
estimates of the average differential velocity
between the thick and the thin disks vary from $0$ to $-30\,\mathrm{km/s}$
(\cite{Beers}, \cite{Binney}).
As long as this constant term is smaller than $\sigma_r$, its
impact on the transverse speed distribution can be neglected.
Otherwise, the microlensing events would be shorter and the
distributions discussed below would be distorted accordingly.
In the case of lensing by a disk object, we neglect the source
proper motion, because of its much larger distance.
\end{itemize}

\begin{figure*}[h!]
\begin{center}
\begin{turn}{0}
\mbox{\epsfig{file=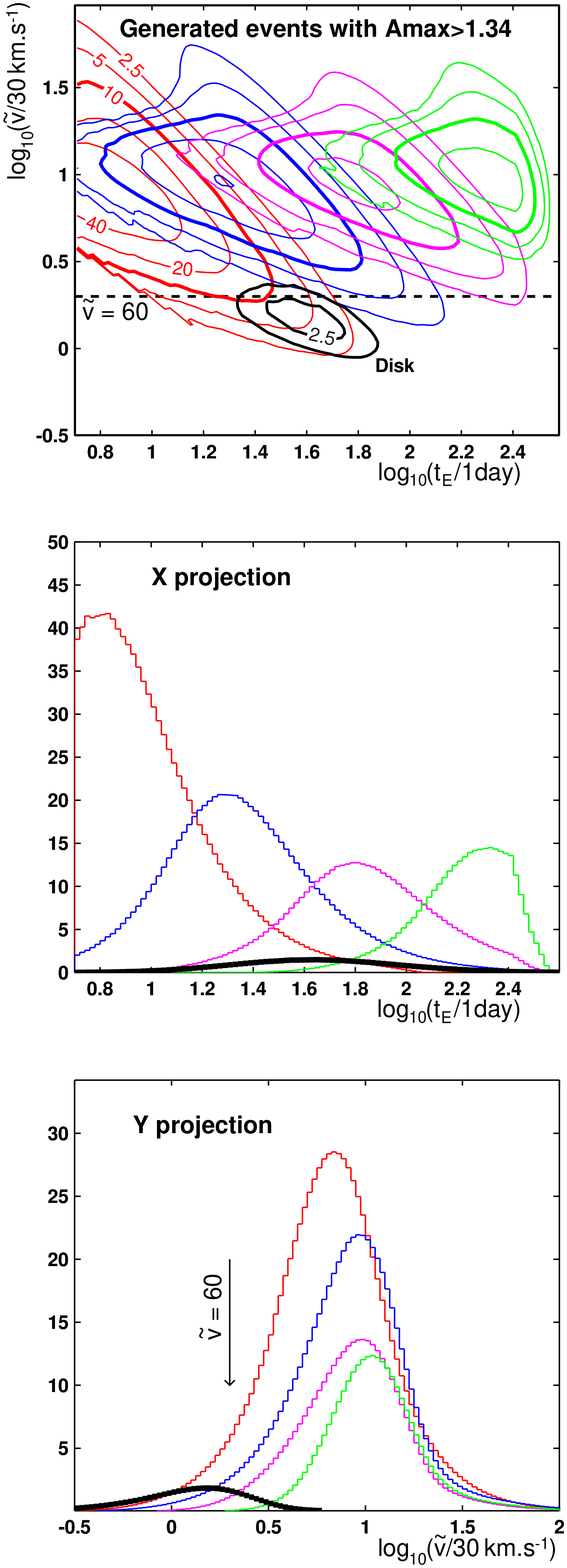,width=6.cm,height=18.cm}}
\mbox{\epsfig{file=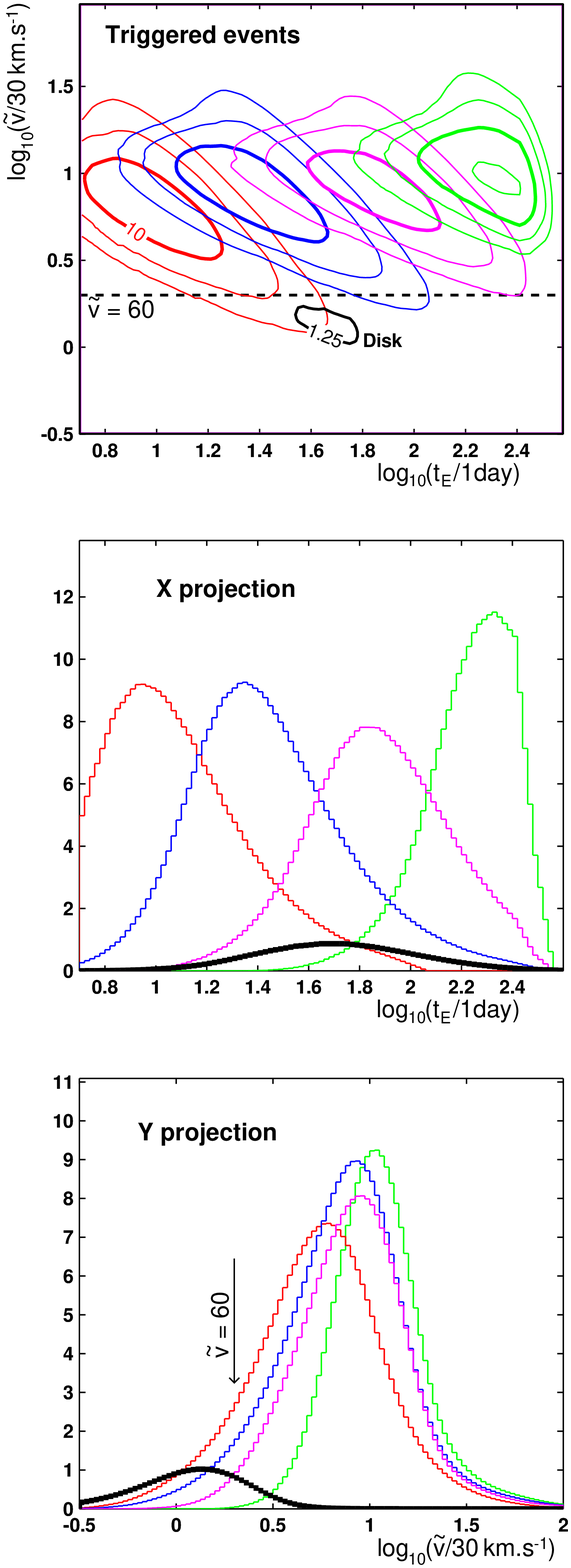,width=6.cm,height=18.cm}}
\mbox{\epsfig{file=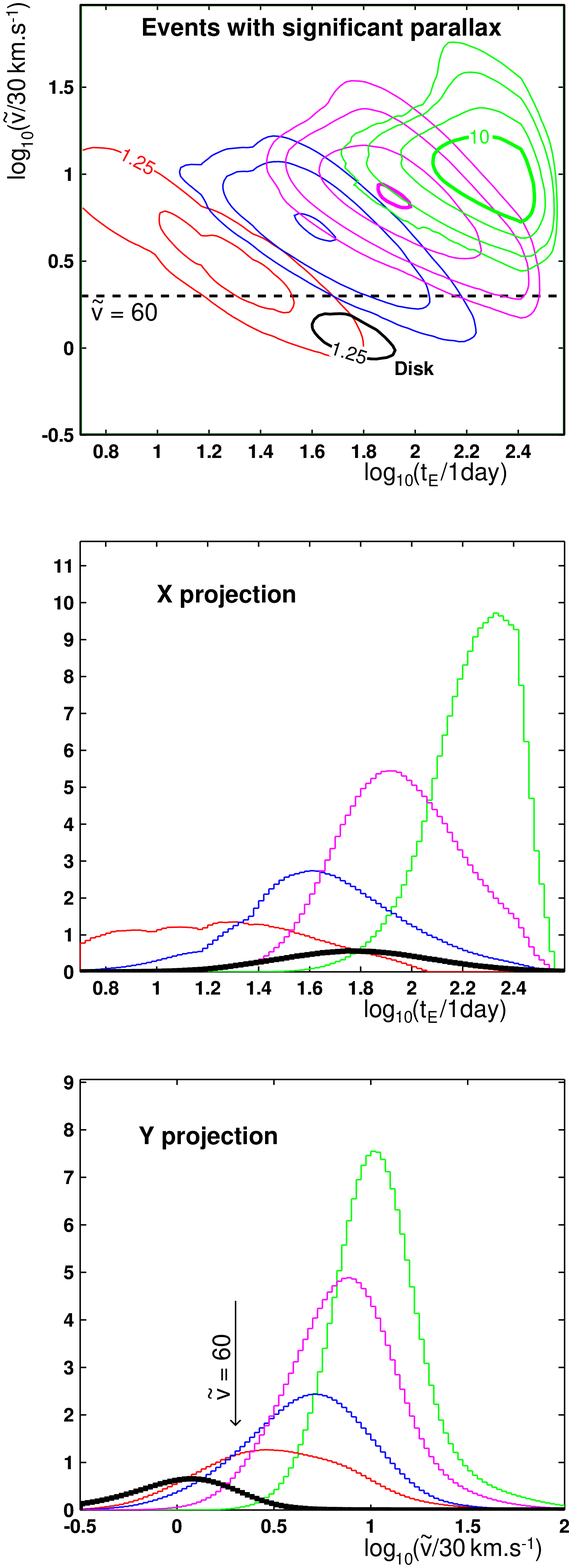,width=6.cm,height=18.cm}}
\end{turn}
\caption[]{Expected distributions of microlensing events
in the $Log_{10}(\tilde v/30\,\mathrm{km.s^{-1}})$ versus $Log_{10}(t_E/1.\,\mathrm{day})$
plane and projections for the various components of model 1.
The maximum contributions of halo objects of $10^{-2}$,
$10^{-1}$, $1.$ and $10.\Msol$, compatible with the observed
microlensing rates, are respectively plotted in each
2D frame from left to right.
The thin disk contribution lies in the lower part of
the 2D frames.\\
%The horizontal lines in the 2D frames correspond to $\tilde v=60\,\mathrm{km.s^{-1}}$.\\
-Left panels: events with $A_{max}>1.34$.\\
-Center: events that satisfy the trigger requirements.\\
-Right: events that exhibits a $3\sigma$ significant parallax.\\
The iso-density contour levels follow a geometrical series: the
difference between consecutive levels of density is a factor 2.
The thick contours always correspond
to 10 events per abscissa unit and per ordinate unit,
for an exposure of $E=1\,\mathrm{year}\times 10^7\,\mathrm{stars}$.
For clarity of the figure, different scales (explicitely indicated)
may be used for the disk contributions.
The vertical scales for the projections give the expected numbers
of events per abscissa unit, for an exposure of $E=1\,\mathrm{year}\times 10^7\,\mathrm{stars}$.
%$Log_{10}(\tilde v/30km.s^{-1})=0.3$ corresponds to $\tilde v=60km/s$.
}
\label{model1}
\end{center}
\end{figure*}

\begin{figure*}[h!]
\begin{center}
\begin{turn}{0}
\mbox{\epsfig{file=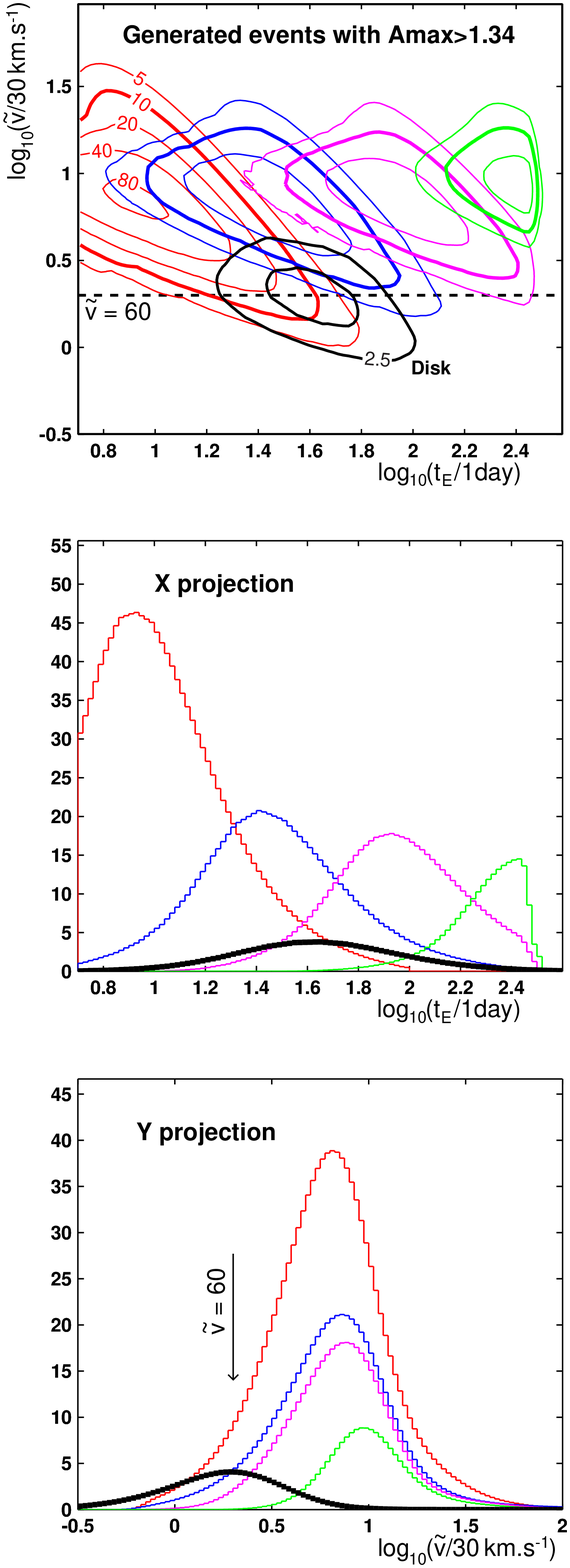,width=6.0cm,height=18.cm}}
\mbox{\epsfig{file=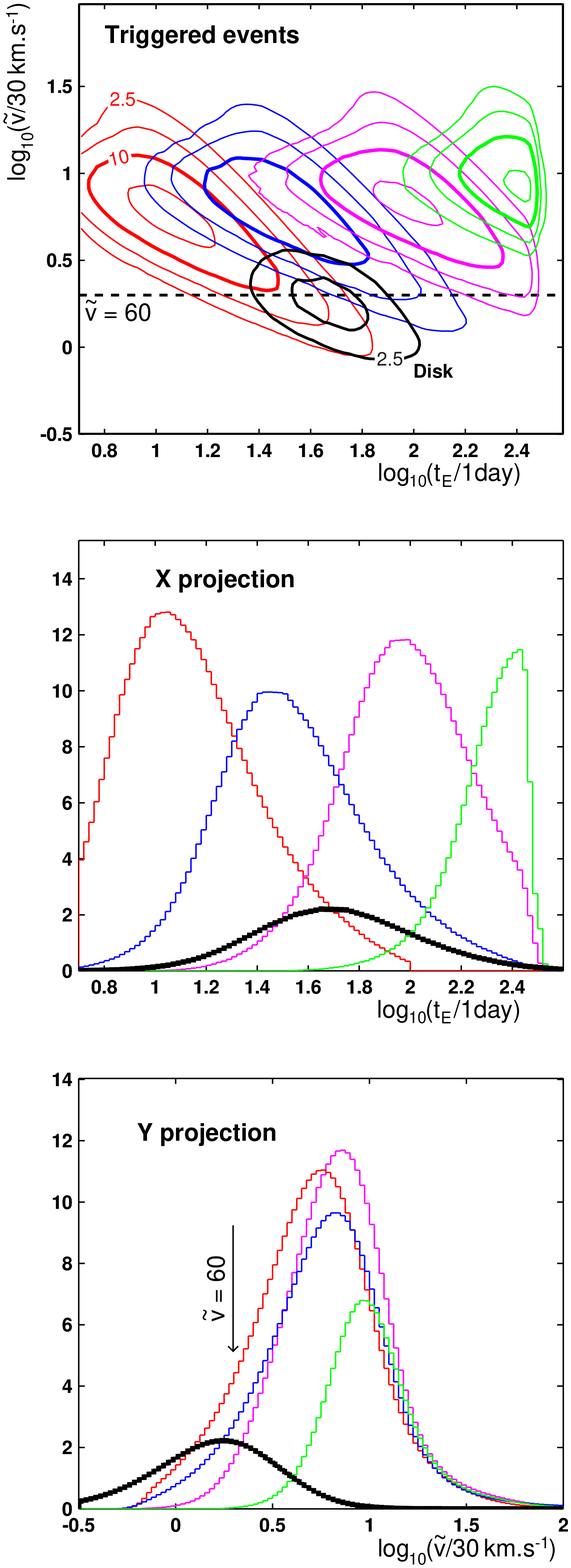,width=6.0cm,height=18.cm}}
\mbox{\epsfig{file=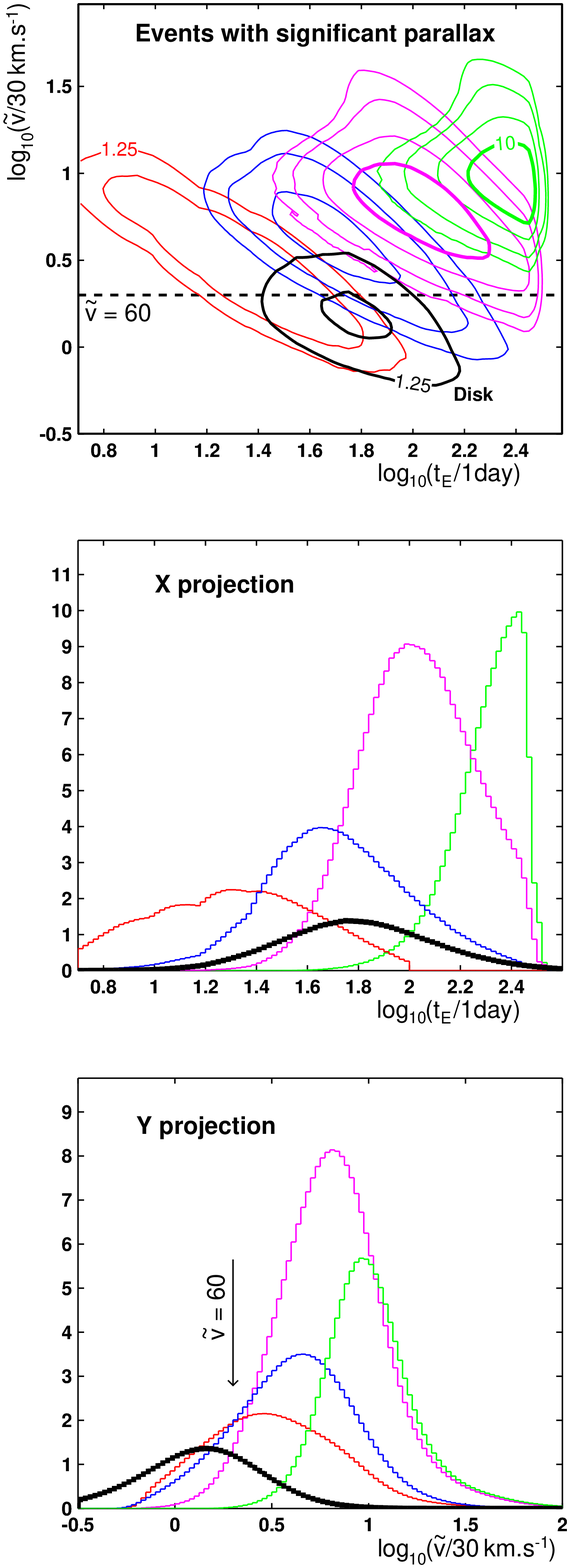,width=6.0cm,height=18.cm}}
\end{turn}
\caption[]{Expected distributions of microlensing events
for model 2 (see caption of Fig. \ref{model1}).
The thin plus thick disk contributions lie in the lower part of the frames.
}
\label{model2}
\end{center}
\end{figure*}

The model parameters we use in this paper are summarized in
Table 2.
\begin{table}
\begin{center}
\caption[]{Parameters of the galactic models used in this article 
and predictions of the rotation curve of the Milky Way.}
\begin{tabular}{|c|l|l|c|c|}
\hline
Structure	& \multicolumn{2}{c|}{Parameter}			&model 1&model 2 \\ \hline
	& \multicolumn{2}{l|}{$\Sigma\ (M_{\odot}\,{\rm pc}^{-2})$}	& \multicolumn{2}{c|}{50} \\ 
	& \multicolumn{2}{l|}{$H\ ({\rm kpc})$}             	& \multicolumn{2}{c|}{0.325} \\ 
Thin	& \multicolumn{2}{l|}{$R_d\ ({\rm kpc})$}		        & \multicolumn{2}{c|}{3.5}  \\
disk	& \multicolumn{2}{l|}{$M_{thin}(\times 10^{10}M_{\odot})$}	& \multicolumn{2}{c|}{4.4} \\
\cline{2-5}
	&velocity	&$\sigma_{r}\ \mathrm{(km/s)}$		& \multicolumn{2}{c|}{34.} \\
	&disper-	&$\sigma_{\theta}\ \mathrm{(km/s)}$	& \multicolumn{2}{c|}{28.} \\
	& sions		&$\sigma_{z}\ \mathrm{(km/s)}$	& \multicolumn{2}{c|}{20.} \\ \hline
\hline
	& \multicolumn{2}{l|}{$\Sigma\ (M_{\odot}\,{\rm pc}^{-2})$} & -    & 30      \\
	& \multicolumn{2}{l|}{$H\ ({\rm kpc})$}	                & -	& 1.0     \\ 
Thick	& \multicolumn{2}{l|}{$R_d\ ({\rm kpc})$}	                & -    & 3.0   \\
disk	& \multicolumn{2}{l|}{$M_{thick}\ (\times 10^{10}M_{\odot})$	}& -	& 2.9 \\
\cline{2-5}
	&velocity	&$\sigma_{r}\ \mathrm{(km/s)}$		& -	& 51. \\
	&disper-	&$\sigma_{\theta}\ \mathrm{(km/s)}$	& -	& 38. \\
	& sions		&$\sigma_{z}\ \mathrm{(km/s)}$	& -	& 35. \\ \hline
\hline
	& \multicolumn{2}{l|}{$\rho_{H\odot}\ (M_{\odot}\,{\rm pc}^{-3})$} & 0.008	& 0.005  \\
Halo	& \multicolumn{2}{l|}{$R_{c}\ ({\rm kpc})$}	& 5.0  & 15.0    \\ 
	& \multicolumn{2}{l|}{$M\ {\rm in}\ 60\,{\rm kpc}\ (10^{10} M_{\odot})$}   & 51	& 22     \\
\cline{2-5}
	 & velocity	&$\sigma$ (km/s)		& 200.	& 200.	\\ \hline
\hline
	& \multicolumn{2}{l|}{$\rho_{\odot}\ (M_{\odot}\,{\rm pc}^{-3})$}	& 0.085	& 0.097	\\
Predictions	& \multicolumn{2}{l|}{$V_{rot}\ {\rm at}\ {\rm sun}\ ({\rm km}\,{\rm s}^{-1})$}   & 192  & 219    \\ 
	& \multicolumn{2}{l|}{$V_{rot}\ {\rm at}\ 50\,{\rm kpc}$}	 & 199  & 182     \\ \hline       
\noalign{\smallskip}
\end{tabular}
\end{center}
\label{tabmodel}
\end{table}
Figures \ref{model1} and \ref{model2} show the expected
distributions of events towards LMC,
for the two Galactic models and the 4 halo mass functions,
respectively peaked at $10^{-2}$,
$10^{-1}$, 1. and $10. \Msol$.
The choice of $\tilde v$ and $t_E$ gives to these distributions
some interesting scaling features because
$\tilde v$ is a projected speed that only depends on the
position and velocity of the deflector (not on its mass),
and $t_E$ scales with $\sqrt{M_{deflector}}$ for a given spatial and
velocity distribution.
Then, for a fixed spatial and velocity distribution
of the compact halo objects, we can predict the scaling of
the $Log_{10}(\tilde v/30\,\mathrm{km.s^{-1}})$ versus $Log_{10}(t_E/1.\,\mathrm{day})$
distribution when changing the mass function, assuming that no
experimental bias is introduced (i.e. before any selection process);
we expect this initial (unbiased)
$Log_{10}(\tilde v/30\,\mathrm{km.s^{-1}})$ versus $Log_{10}(t_E/1.\,\mathrm{day})$
distribution to be simply horizontally translated by +0.5 unit when the mass
of the deflectors is multiplied by 10;
the integral of the distribution
should also be scaled by $1/\sqrt{10}$ times the corresponding fraction
of the halo made of such compact objects.

The left panels of figures \ref{model1} and \ref{model2} reflect these
properties, with limitations due to truncations in $t_E$.
It also appears that the general shape of the halo distribution is almost
unaffected by the choice of a standard or light halo.
The trigger and parallax detection efficiencies
distort considerably these distributions.
\section{Discussion}
We center the discussion on the LMC case, considering
its larger rate of detectable events compared to the SMC case.
Anyway, as already mentioned in Sect. 4, the efficiency diagrams can
also be used as it for the SMC studies.
The halo mass function is mainly bounded
by the upper limit established for each mass
by the microlensing surveys, but its actual shape is unknown.
Then, a given observed microlensing $t_E$ distribution can
generally be explained by the two halo models;
indeed, for each halo, appropriate mass function shapes
(i.e. not anymore simple delta functions)
can be found which, added to the disk(s) contribution(s),
satisfy our published upper limits, and result
in identical microlensing $t_E$ distributions.
Our proposed strategy provides
a sensitivity to the parallax effect that allows to
distinguish the self-lensing
\footnote{A lens located at $x>0.95$ -- case of the self-lensing --
would give high values of $\tilde v=v_T/(1-x)>20\times 30\,\mathrm{km/s}$, and the
distributions for such events are mainly located
in overflow of figures \ref{model1} and \ref{model2}.}
and the halo contributions from
the disk(s) one(s), and also to remove the degeneracy between
our 2 models.
Table 3 gives the rates of events expected to trigger
complementary observations, the rates of events with a $3\sigma$
significant parallax and with $\tilde v<60\,\mathrm{km/s}$, and the
ratios of the parallax events to the trigger events,
assuming 1\% photometric precision after the alert.
As mentioned before, these ratios are unsensitive
to the effective trigger threshold, and can be used with the
effective observed alert rate.
\begin{table}
\begin{center}
\caption[]{Rates of parallax events and ratios to trigger events (in \%),
assuming 1\% photometric precision, for an exposure of $1\,\mathrm{year}\times 10^7\,\mathrm{stars}$.}
\begin{tabular}{|l|c|c|c|}

\hline
Mass of		& Trigger & with detected	& with	\\
halo lenses	& events  & parallax $>3\sigma$	& $\tilde v<60\,\mathrm{km/s}$  \\
\hline
\multicolumn{4}{|c|}{model 1 - standard halo, thin disk}  \\
\hline
$10^{-2} \Msol$	& 4.9	& 0.92 (19\%) & 0.14 (2.8\%)  \\
$10^{-1} \Msol$	& 6.5	& 2.0 (31\%) & 0.28 (4.3\%)  \\
$1. \Msol$	& 5.4	& 3.4 (63\%) & 0.09 (1.7\%)  \\
$10. \Msol$	& 4.9	& 4.0 (82\%) & 0. (0.\%)   \\
\hline
thin disk	& 0.7	& 0.42 (62\%) & 0.36 (53\%)  \\
\hline
\hline
\multicolumn{4}{|c|}{model 2 - light halo, thin+thick disks} \\
\hline
$10^{-2} \Msol$	& 8.3	& 1.9 (23\%) & 0.54 (6.5\%) \\
$10^{-1} \Msol$	& 7.1	& 2.7 (39\%) & 0.44 (6.2\%) \\
$1. \Msol$	& 7.6	& 5.3 (71\%) & 0.16 (2.1\%) \\
$10. \Msol$	& 3.4	& 2.9 (84\%) & 0. (0.\%) \\
\hline
thin+thick disks& 1.7	& 1.0 (59\%) & 0.7 (41\%) \\
\hline
\end{tabular}
\end{center}
\label{ratios}
\end{table}
From this table, we can conclude that requesting for a strong parallax
effect (with $\tilde v<60\,\mathrm{km/s}$) essentially selects disks events, due
to the proximity of their lensing objects.
\begin{itemize}
\item
If the disk contribution to the microlensing signal is dominant
(no compact objects in the halo), then about half of the events
will exhibit a parallax deviation with $\tilde v<60\,\mathrm{km/s}$
\item
If the halo contribution is important, then the fraction of events
with $\tilde v<60\,\mathrm{km/s}$ will be much smaller.
\item
If the so-called ``self-lensing'' contribution is dominant,
then this fraction will be negligible. The self-lensing hypothesis
assumes that the microlensing events are
due to lenses belonging to the LMC or the SMC.
Such events would never give a detectable parallax effect, and our simulation
shows that they would lie above the upper part of the
$Log_{10}(\tilde v/30\,\mathrm{km.s^{-1}})$ versus $Log_{10}(t_E/1.\,\mathrm{day})$ diagram
of figures \ref{model1} and \ref{model2}.
\end{itemize}
\subsection{Distinguishing between the halo models}
\begin{figure}[h!]
\begin{center}
\begin{turn}{0}
\mbox{\epsfig{file=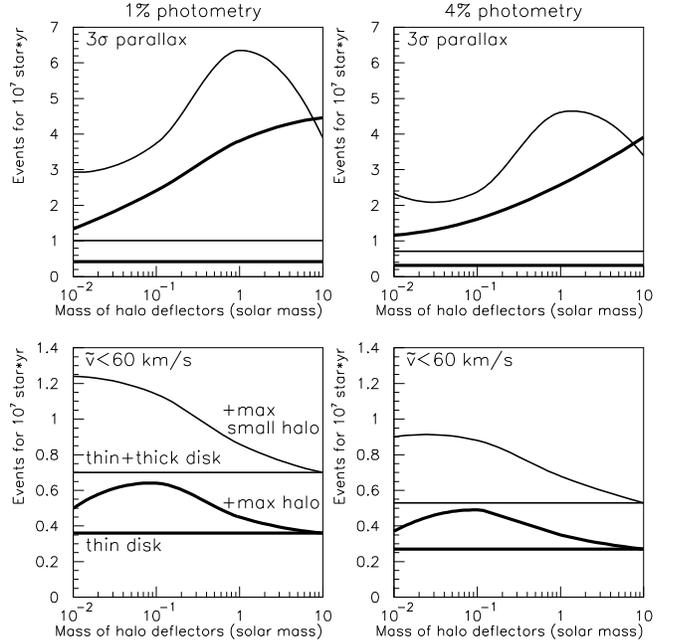,width=9.5cm}}
\end{turn}
\caption[]{
Rates of significant parallax events ($>3\sigma$) (upper panels),
and of events with
$\tilde v<60\,\mathrm{km/s}$ (lower panels) as a function of the halos deflectors mass.
Thick lines correspond to model 1, thin lines to model 2.
The lower (horizontal) lines belong to a null halo contribution,
the upper lines belong to the maximum halo contribution of objects
with mass indicated by the abscissa.
Left panels are calculated for a 1\% photometric precision, right
panels for a 4\% precision.
The numbers are given for one 
year of observation of $10^7$ stars in the field of {\it LMC}.
\label{numbers}}
\end{center}
\end{figure}

\begin{figure}[h!]
\begin{center}
\begin{turn}{0}
\mbox{\epsfig{file=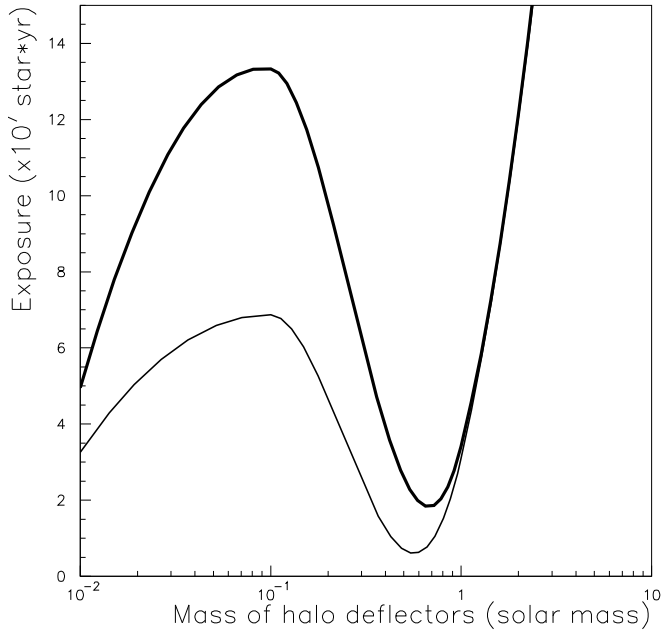,width=9.5cm}}
\end{turn}
\caption[]{
Minimum exposures needed to discriminate between the two halo models
as a function of the mass of the halo objects,
in years for $10^7$ stars of the {\it LMC}.
This exposure corresponds to the same number of years
of parallax studies with an EROS-type alert system.
Notice that this is needed to distinguish between models
with the {\it maximum} allowed fraction of compact objects.
These numbers are based on our simulated trigger efficiency $\epsilon_{trig}$.
As mentioned in the beginning of section 4, they may need
to be normalised if the true efficiency is very different.
The thin line corresponds to 1\% photometric precision, the thick line
belongs to 4\% precision.
\label{exposures}}
\end{center}
\end{figure}

Figure \ref{numbers} gives for the 2 models
the rate of events that exhibit a significant parallax effect ($>3\sigma$),
and the rate of events with a projected speed $\tilde v<60\,\mathrm{km/s}$,
as a function of the halos deflectors mass.
The $\tilde v<60\,\mathrm{km/s}$ cut corresponds to
$Log_{10}(\tilde v/30\,\mathrm{km.s^{-1}})<0.3$ in figures
\ref{model1} and \ref{model2}.
The right part of Fig. \ref{numbers} illustrates the effect of a
photometric precision degradation down to 4\%: the sensitivity
to parallax is typically downgraded by 25\%.
From this figure, one can estimate
the minimal exposure allowing to separate models that
predicts comparable optical depths;
we consider that models 1 and 2 with a {\it maximum} contribution of
halo objects (respectively $\mu_1$ and $\mu_2$ events expected
with $3\sigma$ significance parallax
for $10^7\,\mathrm{star}\times\,\mathrm{year}$), can be distinguished
if the exposure $\alpha\times 10^7\,\mathrm{star}\times\,\mathrm{year}$ is large enough to
allow a significant separation of the two expected distributions
of the events numbers. We choose the convention
that the mean values of the two distributions should be separated by at least
the sum of their widths.
This condition can be expressed as follows:
$$|\alpha\mu_1-\alpha\mu_2|>\sqrt{\alpha\mu_1}+\sqrt{\alpha\mu_2}\ \Longrightarrow
\alpha>\frac{(\sqrt{\mu_1}+\sqrt{\mu_2})^2}{(\mu_1-\mu_2)^2}$$
Figure \ref{exposures} shows the minimum exposure needed to distinguish
between the two models, as a function of the mass of the halo objects.
A few years of parallax studies with an EROS-type alert system
(monitoring $\sim 10^7$ stars) are
needed to separate the two halo models if the halo objects are lighter
than the solar mass. The critical aspect of the photometric precision
is clearly visible on the figure.
A definitive model separation -- taking into account the complete allowed
domains for the mass functions -- will probably need the extraction
of parallax events with $\tilde v<60\,\mathrm{km/s}$, specially if there is very
few halo events (the most pessimistic case: if the halo fraction
made of compact objects is small or if these objects are very heavy).
Then, an exposure of several $10^8\,\mathrm{star}\times\,\mathrm{year}$ is needed,
that should be accessible after
a few years of running a next generation microlensing survey.
Obviously, more sophisticated analysis, such as likelyhood analysis,
should be done after the detection of a reasonable sample
of parallax events.
\subsection{Self-lensing}
If self-lensing is dominant, then very few lenses (only those
belonging to the disk) will produce a detectable parallax.
The relevant parameter to test this hypothesis will be the ratio
of the parallax events to the triggered events.
EROS (phase 2) found 2 events for an exposure of
$3\times 10^7\,\mathrm{star}\times\,\mathrm{year}$
towards the LMC (\cite{notenoughmachos}).
If this rate is due to halo and disk objects then,
according to Table 3, typically 50\% of the
events should have a measurable parallax.
On the other hand, if the observed events are
only due to self-lensing, then none will have a measurable parallax.
About 5 years of parallax studies would then be needed to observe
or not 2 to 3 parallax events that would allow to discriminate between
the two hypothesis\footnote{This is a pessimistic estimate, based on the
partial analysis published so far.}.
\section{Conclusion and perspectives}
In this study, we have investigated a strategy to detect parallax
effects in microlensing with an optimal efficiency,
through the triggering of follow-up observations.
Parallax measurements will allow to distinguish between a nearby
population of lenses (belonging to the disks) and a remote population.
One of the most interesting outcome of such parallax studies
would be to solve the question of
the self-lensing hypothesis, which assumes that the
microlensing events are mostly due to lenses belonging to the LMC or SMC.
A definitive separation between models of lens spatial repartition may need
several years of data taking for a new generation of microlensing
survey.
Such measurements need a high sampling follow-up of the triggered
microlensing events towards LMC and SMC, with a 1\% photometric precision.
A partially dedicated one meter class telescope could achieve this
precision through reasonnably long exposures ($\sim 30\,\mathrm{min.}$).
Finally,
it should be mentioned that such parallax monitoring will also have
other applications. A systematic search for source size effects such
as saturation of the maximum magnification can be performed from
the same data. In favourable cases (small impact parameter events)
this would allow to obtain additional constraints
on the lens configurations, that are complementary to the ones
inferred from parallax analysis.

\end{document}